\documentclass[epj]{svjour}
\usepackage{graphics}
\begin{document}
\title{A Theoretical Study of the Equation of States for Crustal Matter
of Strongly Magnetized Neutron Stars}
\author{Nandini Nag, Sutapa Ghosh and Somenath Chakrabarty$^\dagger$
}                     
\institute{Department of Physics, Visva-Bharati, Santiniketan, India
741 235 \\$^\dagger$E-mail:somenath.chakrabarty@visva-bharati.ac.in }
\date{Received:{\today} / Revised version: date}
\abstract{
We have investigated some of the properties of dense sub-nuclear matter at
the crustal region (both the outer crust and the inner crust region) 
of a magnetar. The relativistic version of Thomas-Fermi (TF) model is
used in presence of strong quantizing magnetic field for the
outer crust matter. The compressed matter in the outer crust, which is
a crystal of metallic iron, is replaced by a regular array of spherically 
symmetric Wigner-Seitz (WS) cells. In the inner crust region, a mixture of 
iron and heavier neutron rich nuclei along with electrons and free neutrons 
has been considered. Conventional Harrison-Wheeler (HW) and Bethe-Baym-Pethick 
(BBP) equation of states are used for the nuclear mass formula. A lot of 
significant changes in the characteristic properties of dense crustal matter, 
both at the outer crust and the inner crust, have been observed. 
}

\maketitle
\section{Introduction}
\label{intro}
Magnetars are the most exotic stellar objects, believed to be strongly
magnetized young neutron stars. The surface magnetic field for such
objects are observed to be $\geq 10^{15}$G \cite{R1,R2,R3,R4}. Then it is 
quite likely that the field at the interior, even at the inner crust region 
may be stronger than the surface value (can be predicted theoretically by 
scalar Virial theorem). If the internal field strength is happened to be so
high, then most of the physical and chemical properties
of dense stellar matter of the magnetars must change significantly from
the conventional picture \cite{R5,R6,R7}. A lot of investigations have already 
been done on the effect of strong quantizing magnetic field on various physical
properties of dense stellar matter inside neutron stars as well as quark
matter inside quark stars or hybrid stars, including the effect on quark-hadron
phase transition at the core region of a compact neutron star \cite{R8}. The 
effect of such strong magnetic field on various elementary processes inside 
neutron stars and quark stars or hybrid stars have also been studied. The 
effect of strong quantizing magnetic field on the $\beta$-equilibration among 
the constituents have also been investigated \cite{R9}. It
has also been shown that strong quantizing magnetic field acts like
a catalyst to generate fermion mass dynamically, i.e., chiral  symmetry
breaking occurs in presence of strong quantizing magnetic field 
\cite{R10,R11,R12}.

In this article we present our investigation on the effect of strong magnetic 
field on the crustal matter of magnetars. The work is divided into two parts:
in the first part, based on one of our very recent work \cite{R13}, we
have investigated the effect of strong quantizing
magnetic field on the outer crust matter. In the second part, we
have studied the properties of compact sub-nuclear matter at the inner crust
region in presence of such strong quantizing magnetic field.

The paper is organized in the following manner: In section 2, 
the effect of strong quantizing magnetic field on the crustal
matter of magnetars is discussed. In the outer crust region, the matter
with dense crystalline structure of metallic iron at sub-nuclear density
are replaced by an array of spherically symmetric WS cells with positively 
charged nuclei at the centre surrounded by non-uniform dense electron gas with
over all charge neutrality. In the inner crust we have used the
conventional HW and BBP equation of states for the nuclear matter
\cite{RR14,R14,R15},
consisting of iron along with some more heavier neutron rich nuclei. In
one of our forth coming article we shall report the effect strong
magnetic field on inner crust matter using some modern type nuclear mass
formulas and make a comparative study with those models.
Here we have presented with detailed numerical computation the effect of strong
magnetic field on the inner crust matter, which as mentioned above is assumed 
to be a mixture of iron, some heavier neutron rich nuclei, which we found to be
specially true in presence of strong quantizing magnetic field,
electrons and free neutrons. The presence of free neutrons are considered  
beyond neutron drip density. Finally, in the last section, we
have given the conclusions and discussed the importance and future
prospects of the present work.

\section{Crustal Matter}
\label{sec:1}
In the first part of this section, we present our study on the effect of
strong magnetic field on the outer crust region with the composition
mentioned in the introduction part. For a 
typical neutron star, the width of the outer crust region is $\sim 0.2-0.4$Km. 
In a recent work we have developed an exact
formalism for the relativistic version of TF model in presence
of strong quantizing magnetic field \cite{R13}. 
This formalism is used within the limitation of TF model \cite{R16}, to obtain 
the equation of state for crustal matter of a typical magnetar. In this model 
it has
been assumed that for a WS cell, the atomic number is $Z$ and the mass number 
is $A$. To make each cell electrically charge neutral,
$Z$ must also be the number of electrons inside the WS cells.

In this model, the modified form of Poisson's equation is given by \cite{R13}
\begin{equation}
\frac{d^2 \phi}{dx^2}=\sum_{\nu=0}^{\nu_{\rm{max}}}
(2-\delta_{\nu 0})(\phi^2(x)-\phi_\nu^2 x^2)^{1/2}
\end{equation}
where $\phi(x)$ is the modified form of electrostatic potential related to the
original Coulomb potential $V(x)$ by the relation
\begin{equation}
\phi(x)=\frac{\mu x}{Ze^2}[\mu_e+eV(x)],
\end{equation}
with $\mu_e$, the electron chemical potential, assumed to be constant
throughout the cell (this is the so called Thomas-Fermi condition), the
dimensionless scaled radial coordinate $x$ is related to the actual radial 
coordinate $r$ of WS cells by the relation: $x=r/\mu$, with 
\begin{equation}
\mu=\left ( \frac{\pi}{2e^3B}\right )^{1/2},
\end{equation}
$e=\vert e\vert$, the magnitude of electron charge, $B$ is the constant
magnetic field, assumed to be along $z$-direction (we have chosen the
gauge $A^\mu\equiv (0,0,xB,0)$),
\begin{equation}
\phi_\nu=\frac{m_\nu\mu}{Ze^2},
\end{equation}
with $m_\nu=(m_e^2+2\nu eB)^{1/2}$, $m_e=0.5$MeV, the electron rest mass
and $\nu=0,1,2,.......,\nu_{\rm{max}}$, is the Landau quantum number for the
electrons, $\nu_{\rm{max}}$ is the upper limit of Landau quantum number 
(the upper limit of the Landau quantum number $\nu_{\rm{max}}$ is finite at 
zero temperature, otherwise $\nu_{\rm{max}}=\infty$).
Finally, the factor $(2-\delta_{\nu 0})$ indicates that the
zeroth Landau level is singly degenerate, whereas, all other states are
doubly degenerate. In this article we have assumed that the electron gas
in the dense crustal matter of crystalline metallic iron is strongly degenerate
and is considered to be at zero temperature. Then it is quite obvious from the
non-negative value of electron Fermi momentum, that the upper limit of 
Landau quantum number $\nu_{\rm{max}}$ is given by
\begin{equation}
\nu_{\rm{max}}=\frac{\mu_e^2-m_e^2}{2eB}
\end{equation}

To obtain numerical solution for $\phi(x)$ for a given magnetic field
and a particular set of $Z$ and $A$, we further assume that instead of a
point object, the nucleus at the centre of a WS cell, has a finite
dimension and is assumed to be spherical in nature, so that the
corresponding radius $r_n=r_0A^{1/3}$, with
$r_0=1.12$fm. Such a choice also removes the singularity problem of TF
equation at the origin \cite{R17}. Therefore it is not necessary to
implement the
prescription given by Feynman, Metropolis and Teller
to obtain the numerical solution for TF equation \cite{R18}. Again from
the physics point of view, the potential must satisfy the boundary conditions, 
given by
\begin{equation}
rV(r)=Ze ~~~ {\rm {for}} ~~~ r\rightarrow r_n, ~~~ {\rm{and}} ~~~
\frac{dV}{dr}=0 ~~~ {\rm{for}} ~~~ r\rightarrow r_s
\end{equation}
where $r_s$ is the radius of the WS cell. Then by simple algebraic manipulation
it is easy to show that $\phi(x)$, the modified form of Coulomb potential
satisfies the boundary conditions
\begin{equation}
\phi(x)\vert_{x=x_n}=1 ~~~ {\rm{and}} ~~~ \frac{d\phi}{dx}
\vert_{x=x_s}=\frac{\phi(x)}{x}\vert_{x=x_s},
\end{equation}
where $x_n=r_n/\mu$, the scaled nuclear radius and $x_s=r_s/\mu$,
the corresponding scaled radius of the WS cell. Both these quantities
are dimensionless.

Further, the right hand side of the Poisson's equation (eqn.(1)) must be
real. Which requires $\phi_\nu x \leq \vert \phi(x)\vert$. This is an
additional condition, to be satisfied by the upper limit of
Landau quantum number, and may be written by the following inequality:
\begin{equation}
\nu_{\rm{max}}(x)\leq \left ( \frac{e^6Z^2}{\pi x^2}\phi(x)^2-\frac{m_e^2}{2eB}
\right ),
\end{equation}
From the definition of Landau quantum number, the above inequality 
must necessarily be $\geq 0$. The above equation also shows that the upper 
limit of Landau quantum number depends on the position ($x$ or $r$ coordinates)
of the electron within the WS cell, with which it is associated.

Since electron distribution is non-uniform within each cell, the Fermi
momentum of a particular electron must depend on its positional
coordinate in the cell. Then it is expected that the variation of
$p_F(r)$ will be such that the electron chemical potential, given by 
\[
\mu_e+ eV(r)=(p_F(r)^2+m_e^2+2\nu(r)eB)^{1/2}
\]
remain constant throughout the cell with proper space dependent Landau
quantum number $\nu$.

Satisfying all these conditions, of which some of them are particularly
necessary for this model, we have solved the Poisson's equation numerically
within the range of $r$ from nuclear surface to the WS cell boundary,
for $B=B_c^{(e)}, 10\times B_c^{(e)}, 10^2\times B_c^{(e)}$ and
$10^3\times B_c^{(e)}$, where $B_c^{(e)}$ is the quantum critical
limit for the magnetic field at and above which the Landau levels are
populated for the relativistic electrons, and may be expressed in terms
of the universal constants, given by
$eB_c^{(e)}=m_e^2$ (with the choice of unit $\hbar=c=1$). The magnetic
field beyond this limit is called the quantizing magnetic field and
the quantum mechanical effect plays an important role in this domain. 
The numerical value of this field is $4.43 \times 10^{13}$G.
Within the range $x_n\leq x \leq x_s$, the numerical solution for $\phi(x)$ 
can be fitted exactly by a straight line for a particular magnetic field
strength $B$ and can be expressed as $\phi(x)=ax+b$. Obviously, the
parameters $a$ and $b$ are functions of magnetic field strength. We have
obtained this functional form by minimizing $\chi^2$ numerically using a
computer code for multi-parameter fitting program.
In Table-I we have shown the variation of the parameters $a$ and $b$ with 
magnetic field strength.

Because of some numerical in-accuracy, the values of the parameter $b$, 
the intersections of the straight lines with $y$-axis, are not exactly 
one. However, for all the magnetic field values,
the parameter $b$ is very close to one. The numerical values for the other 
parameter $a$ is always negative and its magnitude increases with the increase 
in magnetic field $B$. Which actually means that the radius or the volume of a 
particular WS cell decreases with the increase in magnetic
field strength. However, in this particular article, we have avoided the
origin, considering a finite size nucleus of radius $r_n$ at the centre.
Then we have on the nuclear surface $b=1-ar_n/\mu$. In Table-I we have also 
shown the explicit variation of $x_s$, 
the scaled surface radius of the WS cells, for various magnetic field
strength. While solving the Poisson's equation numerically for a given
magnetic field strength $B$, the instruction is given in our
numerical code to terminate if the surface condition, given by eqn.(7) is 
satisfied and hence we obtain $x_s$ (also $r_s=x_s \mu$) as a function of 
magnetic field strength. In fig.(1) we have shown the
variation of $x_s$ and also the corresponding $r_s$ as given in eqn.(9)
below, with the strength of
magnetic field. This figure clearly shows that the WS cells become more
compressed if the magnetic field becomes stronger. This is in some sense
analogous to what is called the {\it{magnetostriction}} in classical
magneto-statics. These two curves can also be fitted by the power law
functions, given by
\begin{eqnarray}
x_s&=&29.43 \times \left (\frac{B}{B_c^{(e)}}\right )^{-0.41}
~~{\rm{MeV}}^{-1}~~
{\rm{and}} ~~~\nonumber  \\
r_s&=&1.704 \times \left (\frac{B}{B_c^{(e)}}\right )^{-0.91}~{\rm{\AA}}
\end{eqnarray}
Knowing the scaled Coulomb potential $\phi(x)$ at various $x$ points (in
the range $x_n$ to $x_s$) within the WS cell for a given magnetic field
strength, we have evaluated numerically $\nu_{\rm{max}}(x)$ at every $x$
points within the cell. The variations of $\nu_{\rm{max}}(x)$ with $x$ for four
different magnetic field strengths are shown in fig.(2). Although 
$\nu_{\rm{max}}(x)$ must be a set of discrete numbers, for the sake of 
illustration, we have plotted it as a continuous variable. In this figure, 
curves $a$ and $b$ are for $B=B_c^{(e)}$ and $10\times B_c^{(e)}$ respectively.
For these two curves, the variables $\nu_{\rm{max}}(x)$ is plotted along the 
left side $y$-axis and $x$ (this is actually $\mu x$ in Mev$^{-1}$) along 
$x$-axis at the bottom. 
Similarly for $B=50\times B_c^{(e)}$ and $100 \times B_c^{(e)}$, the variations
are shown by the curves $\alpha$ and $\beta$ respectively. In this case, 
$\nu_{\rm{max}}(x)$ is plotted along the right side $y$-axis and $x$ is plotted
along upper $x$-axis. From this figure, it is possible to make a number of 
conclusions: (i) For larger $B$ values, $x_s$ are smaller. (ii) For smaller 
$B$, $\nu_{\rm{max}}(x)$ starts with quite large value near the nuclear 
surface, e.g., $=124$ and $=15$ as shown in curves $a$ and
$b$ respectively. Whereas, for curve $\alpha$ it starts with
$\nu_{\rm{max}}=2$ and for $\beta$, the starting value is $\nu_{\rm{max}}=1$.
(iii) The discrete nature of $\nu_{\rm{max}}$ is obvious from the high
magnetic field curves $\alpha$ and $\beta$. (iv) Finally, for all the
field values, the upper limit $\nu_{\rm{max}}(x)$ becomes exactly zero at the
surface of the WS cells. In other wards, we can 
say that all the electrons near the WS cell surface are strongly
polarized and the spins are anti-parallel to the direction of magnetic
field. This is, of course, a purely relativistic effect. The possibility
of fully polarized scenario can easily be obtained from the analytical
solution of Dirac equation for electrons in presence of strong quantizing 
magnetic field. The eigen functions will not be simple spinor solutions,
whereas the energy eigen value will be $E_\nu=(p_z^2+2\nu eB+
m_e^2)^{1/2}$, with $2\nu=n+1+m_s$, where $n=0,1,2, ..$ is the Landau
principal quantum number and $m_s=\pm 1$, the eigen values for the spin
operator $\sigma_z$ \cite{R19,R13}. Hence, for $\nu=\nu_{\rm{max}}=0$, the only
possible choice is the combination $n=0$ and $m_s=-1$. Which actually means 
that in the zeroth Landau level the spins of all the electrons are in the
direction opposite (this is due to negative charge carried by the
electrons) to the external magnetic field. We have further noticed that
beyond the field value $100 \times B_c^{(e)}$ (which is slightly $>10^{15}$G), 
the 
upper limit $\nu_{\rm{max}}(x)$ becomes identically zero not only at the
surface region of WS cells, but at all the points inside the cell.  

To obtain the density distribution of electrons within the WS cells, let
us consider the expression for electron number density, given by
\begin{equation}
n_e(x)=\frac{eB}{2\pi^2}\sum_{\nu=0}^{\nu_{\rm{max}}(x)} (2-\delta_{\nu 0})
p_F(x)
\end{equation}
where the electron Fermi momentum $p_F(x)$ can be obtained from the TF
condition and is given by
\begin{equation}
p_F(x)=\left \{ Z^2e^4\left (\frac{\phi(x)}{\mu x}\right )^2-m_\nu^2\right
\}^{1/2}
\end{equation}

Since the electron density is larger near the nuclear surface inside the WS 
cells compared to the boundary values,
the effect of electron density dominates over the influence of 
magnetic field strength at the central region, whereas near the WS surface, the
magnetic field will play significant role. This will naturally make the upper
limit $\nu_{\rm{max}}$ large enough (except for $B \geq 10^{15}$G) near
the nuclear surface and quite small or identically zero near WS cell
boundary. The results are also quite obvious because of the electron Fermi
momentum, which  is minimum at the nuclear surface and maximum near the the WS
cell boundary. Later we shall show that the electron kinetic energy
will also behave like Fermi momentum.
Therefore. the quantum mechanical effect for low and moderate values magnetic 
field strength will be more significant near the surface region compared
to the core region, where the effect is almost classical. This is
particularly obvious from the curves $a$ and $b$ of fig.(2). For
these curves, $\nu_{\rm{max}}(x)$ is naturally started with quite large values.

Using the numerically fitted form of $\phi(x)$ we obtain $p_F(x)$ for a
given magnetic field strength and the corresponding $n_e(x)$.  In fig.(3) we 
have plotted the variation of electron density 
within WS cells for four different magnetic field strengths. In this figure,
the curves indicated by the numbers $0,1,2$ and $3$ are for
$B=B_c^{(e)}$, $B=10\times B_c^{(e)}$, $B=10^2\times B_c^{(e)}$ and 
$B=10^3\times B_c^{(e)}$ respectively. The physical reason behind such
increase in electron density is because of the decrease in WS cell
volume with the increase in  magnetic field strength, whereas, the total
amount charge within the cell remain constant. It is quite obvious from these 
curves that the electron density increases with the increase in magnetic field 
strength. Further, for all the values of $B$, electron density is maximum at 
the centre and minimum at the surface. The oscillatory nature of $n_e(x)$ is 
because of the crossing of various Fermi levels within the cell. This is
a kind of de Haas van Alphen effect.
In our model we have assumed eqn.(10) as given in \cite{R13}. One must
use the condition \cite{LY}
\begin{equation}
\epsilon_e(r)-eV(r)+E_{ex}(r)=\mu_e=~~{\rm{constant}}~~
\end{equation}
then re-cast the differential equation as given by eqn.(1) in this
article. Here $E_{ex}$ is the electron-electron exchange energy
per particle. However, with this choice it will be an almost impossible task
to solve the problem even numerically. Hence we have made approximation by 
using eqn.(10) of \cite{R13}.

Let us now evaluate the variation of (i) electron kinetic energy, (ii)
electron-nucleus interaction energy, (iii) electron-electron
direct interaction energy and (iv) electron-electron exchange
energy within the WS cells. The kinetic energy part for
electrons at a particular point ($r$) within the WS cell is given by 
\begin{eqnarray}
E_{KE}(x)&=&\int_{r_n}^{r} d^3r
\frac{eB}{2\pi^2}\sum_{\nu=0}^{\nu_{\rm{max}}(r)}
(2-\delta_{\nu 0})\nonumber \\ &&
\int_0^{p_F(r)} dp_z [(p_z^2+m_\nu^2)^{1/2}-m_e]
\end{eqnarray}
Evaluating the integral over $p_z$ analytically and then substituting for
$p_F(x)$ from eqn.(11), with the numerically fitted functional form of 
$\phi(x)$, one can obtain 
the electron kinetic energy $E_{KE}$, as a function of $r$ or $x$
by the numerically evaluating the above integral.
We have seen that within the WS cells, for constant $B$, the 
kinetic energy part satisfies a power law of the form $E_{KE}=\alpha x^\beta$, 
where the parameters $\alpha$ and $\beta$ are functions of
magnetic field strength.  In Table-I we have shown these variations. 
The change is more significant for $\alpha$
than $\beta$. Therefore, just like the electron density, the electron
kinetic energy is also an increasing function of radial coordinate $x$ 
but does not show oscillation. 

Next we consider the three possible types of interaction potential. Let us 
first consider the electron-nucleus interaction part, given by \cite{R13}
\begin{eqnarray}
&&E_{en}(r)=-Ze^2\int_{r_n}^{r}d^3r \frac{n_e}{r}\nonumber \\
&=&-4\pi Z e^2\mu^2\int_{x_n}^{x} xdx n_e(x)
\end{eqnarray}
To obtain this quantity within the WS cell for a 
constant $B$, we substitute the expression for electron number density $n_e(x)$ 
from eqn.(10), and then numerically integrate over $x$ In fig.(4) we have 
shown the variation of $-E_{en}(x)$ (which is a positive quantity) as a 
function of $x$
within the WS cell for three different magnetic field strengths:
$B=B_c^{(e)}, 10^2\times B_c^{(e)}$ and $10^3\times B_c^{(e)}$. 
For very low $x$ values, i.e., almost on the surface of the nucleus at
the centre, the magnitude of the potential energy is an increasing function of 
$x$, which actually means that the region is strongly
attractive in nature. Again at the surface region, just at the skin of the
cell, it is again an increasing function of $x$ (with larger gradient),
so that the attractive field again becomes extremely
strong at the surface region to keep the outer most electrons confined within 
the cell.  At the middle region, there is a kind of saturation and force due to
electron-nucleus interaction almost vanishes. This is quite analogous to
the normal metallic scenario, in which electric field can not exist.

Next we consider the electron-electron interaction. Let us first
evaluate the direct term. It is given by \cite{R13}
\begin{equation}
E_{ee}^{(d)}=\frac{1}{2}e^2\int d^3r n_e(r)\int d^3r^\prime n_e(r^\prime)
\frac{1}{\vert \vec r-\vec {r^\prime} \vert}
\end{equation}
Assuming $\vec r$ as the principal axis and $\theta$ is the angle between
$\vec r$ and $\vec {r^\prime}$, we have $d^3r~=~4\pi~ r^2~~ dr$,\\ 
$d^3r^\prime= 2\pi
{r^\prime}^2dr^\prime \sin \theta d\theta$ (we have assumed that the vectors 
$\vec r$ and $\vec {r^\prime}$ are on the same plane) and $\vert \vec r-\vec
{r^\prime}\vert = (r^2+{r^\prime}^2-2rr^\prime \cos \theta)^{1/2}$. The limits
for both $r$ and  $r^\prime$ are from $r_n$ to $r_s$ and the range of 
$\theta$ is from $0$ to $\pi$.
Then evaluating the trivial integral over $\theta$  the  direct interaction, 
part reduces to the following simple form:
\begin{eqnarray}
E_{ee}^{(d)}&=&8e^2\pi^2 \Big \{\int_{r_n}^{r}rdr n_e(r)
\int_{r_n}^r{r^\prime}^2dr^\prime n_e(r^\prime)\nonumber \\
&+& 
\int_{r_n}^{r}r^2dr n_e(r)
\int_r^{r_s}r^\prime dr^\prime n_e(r^\prime)\Big \}
\end{eqnarray}
Following the same procedure adopted for $E_{en}$ case, we have evaluated
numerically the above coupled integrals. In fig.(5) we have shown the variation
of $E_{ee}^{d}(x)$ within the WS cell for three different magnetic field 
strengths: $B=B_c^{(e)}, 10^2\times B_c^{(e)}$ and $10^3\times B_c^{(e)}$.
Surprisingly, the variations are almost identical with $\vert E_{en}
\vert $. However, the magnitudes are several orders less than the
corresponding electron-nucleus part. Of course, unlike the $E_{en}$,
the direct term is positive throughout and for all the values of magnetic field
strength. 
The same kind of qualitative nature of $\mid E_{en}\mid$ and $E_{ee}^{(d)}$
(figs.(4)-(5)) makes the system of electron gas within the WS cell energetically 
stable. Of course the kinetic energy part and the exchange part also play 
important role. The combination of kinetic energy and three interaction 
potentials of which electron-nucleus interaction part and the exchange part are
negative, gives a constant energy for a particular electron at any point within 
the WS cell. This constant energy is the electron chemical potential and is also 
the minimum possible energy at a particular point.

Next we shall consider the exchange term, which is negative in nature. 
The magnitude of exchange energy integral 
corresponding to the $i$th. electron in the cell is given by \cite{R13}
\begin{equation}
E_{ee}^{(ex)}= \frac{e^2}{2}\sum_j \int d^3r d^3r^\prime \frac{1}{\vert \vec
r-\vec {r^\prime} \vert} \bar\psi_i(\vec r)\bar\psi_j(\vec {r^\prime})
\psi_j(\vec r)\psi_i(\vec {r^\prime})
\end{equation}
where the spinor wave function $\psi(\vec r)$ is given by eqns.(2)-(5)
in \cite{R13} and $\bar \psi(\vec r)=
\psi^\dagger(\vec r)\gamma_0$, the adjoint of the spinor and $\gamma_0$ is
the zeroth part of the Dirac gamma matrices $\gamma_\mu$. Now it is very
easy to show that for $t=t^\prime$
\begin{eqnarray}
&&\bar \psi_i(\vec r)\psi_i(\vec{r^\prime}) = \frac{2m}{L_yL_zE_\nu}
\exp[-i\{ p_y(y-y^\prime)+p_z(z-z^\prime)\}]\nonumber \\ &&
\{I_{\nu;p_y}(x)I_{\nu;p_y}(x^\prime)+I_{\nu-1;p_y}(x) I_{\nu-1;p_y}(x^\prime)\}
\end{eqnarray}
Similarly, we have
\begin{eqnarray}
&&\bar \psi_j(\vec {r^\prime})\psi_j(\vec r) = \frac{2m}{L_yL_zE_\nu^\prime}
\exp[i\{ p_y^\prime(y-y^\prime)+p_z^\prime(z-z^\prime)\}]\nonumber \\ &&
\{I_{\nu^\prime;p_y^\prime}(x)I_{\nu^\prime;p_y^\prime}(x^\prime)+
I_{\nu^\prime-1;p_y^\prime}(x) I_{\nu^\prime-1;p_y^\prime}(x^\prime)\}
\end{eqnarray}
where $I_{\nu ; p_y}(x)$ is same as $I_{\nu}$, given by eqn.(5) in \cite{R13}.
When these two terms are combined, we have, after replacing the sum over $j$
by the integrals
\[
L_yL_z\int_{-\infty}^{+\infty} dp_y^\prime
\int_{-p_F}^{+p_F} dp_z^\prime
\]
\begin{eqnarray}
&&E_{ee}^{(ex)}=\left (\frac{e^2}{2}\right )\left (
\frac{4m^2}{L_y^2L_z^2 E_\nu}\right )
\sum_{\nu^\prime=0}^{\nu_{\rm{max}}}(2-\delta_{\nu^\prime 0})\nonumber
\\ && \int...\int
L_ydp_y^\prime L_zdp_z^\prime d^3rd^3r^\prime\frac{1}{E_{\nu^\prime}}
\frac{1} {\vert \vec r -\vec {r^\prime}\vert} \nonumber \\ &&
\exp[-i\{ (p_y-p_y^\prime)(y-y^\prime)+(p_z-p_z^\prime)(z-z^\prime)\}]
\nonumber \\ && [\{I_{\nu;p_y}(x)I_{\nu;p_y}(x^\prime)+I_{\nu-1;p_y}(x)
I_{\nu-1;p_y}(x^\prime)\} \nonumber \\
&& \{I_{\nu^\prime;p_y^\prime}(x)I_{\nu^\prime;p_y^\prime}(x^\prime)+
I_{\nu^\prime-1;p_y^\prime}(x) I_{\nu^\prime-1;p_y^\prime}(x^\prime)\}]
\end{eqnarray}
It is possible to evaluate the integrals over $y^\prime$ and $z^\prime$, given 
by \cite{R20} 
\begin{eqnarray}
&&\int_{-\infty}^{+\infty} \int_{-\infty}^{+\infty} dy^\prime dz^\prime 
\frac{1} {\vert \vec r -\vec {r^\prime}\vert}\nonumber \\
&& \exp[-i\{(p_y-p_{y^\prime})(y-y^\prime)+ (p_z-p_{z^\prime})(z-z^\prime) \} ]
\nonumber \\ &=& \frac{4\pi}{2K}\exp(-K\vert x-x^\prime\vert)
\end{eqnarray}
where $K=[(p_y-p_{y^\prime})^2+(p_z-p_{z^\prime})^2]^{1/2}$. Further, the
integral over $y$ and $z$ is given by 
\[
\int_{-\infty}^{+\infty} \int_{-\infty}^{+\infty} dy dz=L_yL_z
\] 
Since the final form of exchange integral is multidimensional in nature
with an extremely complicated structure 
of integrand, it is absolutely impossible to have analytical solution
for $e-e$ exchange potential. We have therefore adopted the multi-dimensional
Monte-Carlo integration code to evaluate the exchange term.
It is found that the variation of the magnitude of exchange energy with the scaled 
radius $x$ within the WS cells for a constant $B$ can be expressed by the power
law formula, given by
\[
\mid E_{ee}^{(\rm{ex})}\mid=p x^q,
\]
where the parameters $p$ and $q$ are again
functions of magnetic field strength $B$. The variation of the parameters 
$p$ and $q$ with magnetic field are shown in Table-I.

From the variation of the parameters $p$ and $q$ with $B$, it is quite
clear that the magnitude of exchange energy also increases with $x$ and
also with the magnetic field strength $B$, i.e., minimum at
the central nuclear surface and maximum at the WS cell boundary.

The overall pressure term can also be obtained for the electron gas from
the total energy $E_{\rm{tot}}$, given by
\begin{equation}
P(x)=n_e^2(x)\frac{\partial E_{tot}}{\partial n_e}=P_k(x)+P_c(x)
\end{equation}
where 
\begin{equation}
P_k=\frac{eB}{2\pi^2}\sum_{\nu=0}^{\nu_{\rm{max}}} (2-\delta_{\nu 0})
\int_0^{p_F}\frac{p_z^2} {(p_z^2+m_\nu^2)^{1/2}} dp_z,
\end{equation}
is the electron kinetic pressure part.
The momentum integral for the kinetic pressure can very easily be
obtained, and is given by 
\begin{eqnarray}
P_k&=&\frac{eB}{2\pi^2}\sum_{\nu=0}^{\nu_{\rm{max}}}(2-\delta_{\nu 0})
\Big [
p_F(p_F^2+m_\nu^2)^{1/2}- \nonumber \\
&&m_\nu^2 \ln \left (\frac{p_F+(p_F^2+m_\nu^2)^{1/2}}
{m_\nu} \right ) \Big ]
\end{eqnarray}
Since $p_F=p_F(x)$, this expression will give the electron kinetic pressure
at various points within the WS cell. The other term, $P_c$, coming from
the interaction part, $E_c=E_{en}+E_{ee}^{(d)}+E_{ee}^{(ex)}$, 
and is given by
\begin{equation}
P_c(x)=n_e^2(x)\frac{\partial E_c}{\partial n_e},
\end{equation}
The interaction part of electron pressure is obtained numerically by
substituting the expression for $p_F(x)$ and evaluating the derivatives
over $n_e$ numerically for a given $B$, at a particular 
point $x$ within the cell and also for a
given set of $A$ and $Z$. We have noticed that at the central region of WS
cells, since the density dominates over the magnetic field part, the
total pressure is positive, while at the outer end, the surface region, since 
magnetic field dominates over the density contribution, it becomes negative.
It is found numerically that at some point within the WS cell for a given $B$, 
the total pressure
becomes exactly zero. It is found that this critical position is a function of 
magnetic field strength and decreases with the increase in magnetic field
strength. In fig.(6) we have plotted this variation. The functional
dependence can also be expressed as 
\begin{equation}
x_c=\alpha+\beta\exp\left [ -\gamma \frac{B}{B_c^{(e)}}\right ]
~~{\rm{MeV}}^{-1}
\end{equation}
with $\alpha=0.38$, $\beta=1.28$ and $\gamma=0.171$. If the quantum mechanical 
effect of strong magnetic field dominates over the density effect, electrons 
either occupy only the zeroth Landau level or low lying levels. Which further 
causes a decrease in pressure term. As the magnetic field becomes strong enough, 
the density effect vanishes, even near the centre, which makes $x_c$ extremely 
small. 

In the second part of this section we shall study the effect of strong quantizing 
magnetic
field on inner crust matter at sub-nuclear density.
In the conventional neutron star model, for a typical neutron star of radius 
$\approx 10$km, the width of inner
crust is about $0.5-0.6$km. 
It is also believed that the free neutrons in the inner crust
region may be in super-fluid state. The density range of this zone
depends on the type of equation of state or the mass formula for the
heavy nuclei, including the highly neutron rich nuclei. The density range
in the conventional picture is $\sim 10^9-10^{12}$g/cm$^3$. To investigate the 
properties of inner crust matter, here also we replace the metallic iron atoms
and other heavier atoms with neutron rich nuclei, by WS cells as has been done 
for the outer crust matter. We further assume that the free neutrons in this
region, above the neutron drip point are in normal fluid state.

In this part, to study the effect of strong quantizing magnetic field
on inner crust matter, we consider two types of conventional nuclear
mass formulae, which are generally used below the nuclear saturation
density, but applicable near neutron
drip region. These two mass formulae are (i) Harrison-Wheeler (HW) equation of 
state, and (ii) Bethe-Baym-Pethick (BBP) equation of state
\cite{RR14,R14,R15}. We have already mentioned that in a future communication
the effect of strong magnetic field on the equation of states with
more modern type nuclear mass formulas will be reported.

In HW equation of state, the relativistic electrons are assumed to be in
$\beta$-equilibrium with the nuclei (which also includes the neutron rich
nuclei). Whereas, above the neutron drip density, the $\beta$-equilibrium is 
among the nuclei, free neutrons and the electrons. We have noticed that the
presence of a strong quantizing back ground magnetic field
modifies the  $\beta$-equilibrium configuration significantly although we have
assumed that the electrons are only affected by the presence of strong magnetic
field. As a consequence 
of which more heavier neutron rich nuclei are produced in presence of strong
magnetic field in the inner crust region.
In the conventional astrophysical scenario, in 
presence of free electron gas in stellar medium, the balance between Coulomb 
force and the nuclear force, which gives $^{56}Fe$ as the most stable nucleus
will get shifted towards the heavier nuclei. The nuclei, formed in such an 
environment will contain more neutrons (by inverse $\beta$-decay) compared to 
the usual picture. The Coulomb force plays a very little role. The nuclei 
becomes more and more neutron rich as the electron density goes up and when the
matter density increases to $\sim 4\times 10^{11}$g/cm$^3$, the ratio $n/p$ 
reaches a critical
level. Any further increase in density will lead to neutron drip in the
medium. In such a scenario, highly neutron rich nuclei, electrons and
free neutrons co-exist in chemical ($\beta$) equilibrium. With the further
increase in matter density, beyond neutron drip, nuclei, including the heavier
ones melt and free neutrons
appear in the medium and at some stage the kinetic pressure of the system will 
be dominated by the free neutron pressure instead of electron pressure which may
become negative if the magnetic field is strong enough. 
We shall see later that this positive
pressure contribution of free neutrons make the total pressure
within the inner crust region a positive definite, even in presence of strong
magnetic field. Whereas in the outer crust, the total pressure, coming
from the electron gas only and is negative
at the surface region of WS cells in presence of strong magnetic field.

The BBP equation of state is applicable for nuclear matter in the
density range from neutron drip density $\rho_{\rm{drip}}$ to normal
nuclear density $\rho_{\rm{nuc.}}$ ($\sim 2.8\times 10^{14}$g cm$^{-3}$). 
It is well known
that BBP equation of state is a considerable improvement over the HW
equation of state. In BBP equation of state, a mass formula, more or
less like the HW equation of state is used but incorporated a lot of
improvements from detailed many body calculation. The nuclear surface
energy, for example, assumed in the previous treatment to be that of a
nucleus in vacuum. An introduction of free neutron gas out side the
nuclei reduces the nuclear surface energy. This is quite correct,
because when inside and out side of a nuclei become identical, the
surface energy must vanish. In BBP equation of state, the nuclear Coulomb
energy is included in a more accurate manner and is called {\it{nuclear
lattice Coulomb energy}}. It is well known that the BBP equation of state is 
applicable up to nuclear density, therefore, when all the bound nuclei dissolves
into a continuous matter, mainly composed of free neutrons and a tiny fraction 
of protons and electrons, to incorporate this type of melting process of 
nuclei, in the BBP equation of state the factors giving the fractional
volume occupied by the nuclei and the fractional volume occupied by
the free neutron gas have been taken into account. It is also assumed that at 
this density ($< \rho_{\rm{nuc.}}$) the nuclei
are stable to $\beta$-decay. Further, the neutrons in the free neutron
gas are in chemical equilibrium with the electrons and the nucleons
within the nuclei and in addition to the $\beta$-stability of the
nuclei, the whole system must be in chemical equilibrium. In this model the 
kinetic pressure of free neutrons must necessarily be equal with the pressure 
of the nucleons bound within the
nuclei. This is the condition for mechanical equilibrium. 
Since detailed mathematical formalism for both HW and BBP equation of
states, which are assumed to be not affected by strong magnetic fields,
are available in a large number of classic papers, including the
original ones, and also included in many standard text books
\cite{RR14,R14,R15}.
The effect of strong quantizing magnetic field on the electron part has
already been discussed in the previous section. In this
section, we therefore present only the numerical results  
based on these two equation of states along with the results related to
electron gas where ever needed, as developed in the first part of this section.

In the numerical computation, we have found that for very low mass number, 
since there is no free neutrons available in the system, even much above the 
neutron drip density, only
electrons contribute in total kinetic pressure and is negative if the magnetic 
field affects the electron part quantum mechanically. This
result is almost identical with the outer crust scenario.
In fig.(7) we have plotted the variation of the critical value of
mass number $A$ and the corresponding atomic number $Z$ with the
strength of magnetic field using HW and BBP equation of states, 
These are the critical values at which the kinetic pressure of the inner
crust region just becomes zero, i.e., the system just becomes
mechanically stable. 
It is obvious from this figure that the minimum of mass number of the
nuclei for which the inner crust matter just becomes mechanically stable,
increases with the strength of magnetic field. In other words, this is
equivalent to say that the
presence of strong magnetic field makes the nuclei more and more
massive or increases the minimum mass of the nuclei in the inner crust region. 
In some of our previous work, done long ago \cite{R12}, we have seen that the 
same conclusion is applicable for quark matter. The
presence of strong quantizing magnetic field generates quark mass
dynamically in dense quark matter composed of massless quarks. This is the well
known magnetic field induced chiral symmetry violation. The strong quantizing 
magnetic field acts like a catalyst to generate / increase mass. 
In the BBP equation of state the
minimum mass of the nuclei are more than that obtained from HW equation
of state. Further, one should notice from this figure (figs.(7)) that for 
low and moderate strength of magnetic field, the
effect is not so significant. But beyond $10^{15}$G, when all the electrons 
occupy the zeroth Landau level, or in other words, when the quantum mechanical
effect of strong magnetic field is most
important, the minimum mass rises sharply to very large values. One can
see from \cite{R12} that the
qualitative nature of the curve showing the dependence of dynamical
quark mass on the magnetic field strength is exactly identical, although
the physical scenarios are completely different.

In fig.(8), we have plotted the variation of the ratio $n_e/n$ and
$n_n/n$ with the mass number using HW equation of state. Here $n_e$ is
the electron density, $n_n$ is the free neutron density and
$n=n_n+n_e A/Z$ is the total baryon density of inner crust matter.
Dashed curve is for $B=0$, middle and the lower curves, indicated by
$e1$ and $e2$ are for
$B=10^2B_c^{(e)}$ and $10^3B_c^{(e)}$ respectively. The curve indicated
by $n$ is for free neutron gas. In the HW equation of state, the neutron
number density above the neutron drip point is independent of magnetic field
strength. Further, the
mass number around which neutrons are liberated from neutron rich nuclei
is about $95$, which is again independent of magnetic field strength.
However, immediately after the emission of neutrons from heavy neutron
rich nuclei the overall kinetic pressure can not become non-zero.
In fig.(9), we have plotted the same kind of variations as shown in
fig.(8). Here we have used the BBP equation of state.
Solid curves are for $B=0$, while the dashed curves are for
$10^3B_c^{(e)}$. In this case the neutron drip out from heavy
neutron rich nuclei around $A=100$, which is little bit heavier and more 
neutron rich than
the HW case. Further, the qualitative nature of $x_n=n_n/n$ is totally
different from HW case. Instead of increase initially and then
saturates, as we observe in HW case, it decreases monotonically with A.
Also, the free neutron density depends on the strength of
magnetic field. However, the dependence is not so significant. 
These qualitative differences are found to be the
consequence of chemical equilibrium among the free neutrons, electrons
and nucleons within the heavy neutron rich nuclei and the
overall $\beta$-equilibrium condition.

In fig.(10). the variation of total nuclear density is plotted against the
mass number for both  HW and BBP equation of states. For the HW case, four 
different curves with magnetic field
strengths: $B=0$, $B=10B_c^{(e)}$, $B=10^2B_c^{(e)}$ and
$B=10^3B_c^{(e)}$ are indicated in the diagram by $0$, $1$, $10^2$ and
$10^3$ respectively and shown by solid lines. 
In HW model, initially for low $A$ values, only the
bound nucleons within the nuclei ($=An_e/Z$) contributes. Just beyond
the critical value, $A=95$, since neutrons drip out from the heavy
neutron rich nuclei, the number density jumps suddenly. Of course, below
$A=60$, we can not have stable inner crust matter.
The same kind of variation
for BBP equation of state are shown by the dashed curves and indicated by 
the symbols $a$, $b$ and $c$. 
In this case,
again
because of chemical equilibrium conditions, the
qualitative and the quantitative nature of the curves are 
completely different.

The equation of state from both HW and BBP mass formulas are plotted in 
fig.(11). 
From these two figures it is quite obvious that in presence of strong
quantizing magnetic field, the inner crust matter becomes mechanically
stable only at very high density, when enough number of free neutrons
are available in the system. In this high density situation positive neutron
kinetic pressure dominates over negative electron
pressure in presence of strong quantizing magnetic field. This
is true for both HW and BBP equation of states. Further, it is also
obvious from the figures, that the matter becomes softer with the
increase in magnetic field strength. The qualitative nature of both the
equation of states are almost identical in presence of strong magnetic
field. However, the HW equation of state is a bit softer than the BBP
case. 

\section{Conclusions}
\label{sec:4}
In the first part of this article we have presented our investigation on
the properties of dense outer crust matter of the magnetars. We have
replaced the dense metallic iron crystal by a regular array of
spherically symmetric WS cells. It has been observed that the radius of
each cell decreases with the magnetic field strength. In this article,
however, we
have not considered the magnetic field induced deformation of WS cells,
which is quite likely if the magnetic field is strong enough. In
future we shall present this important issue with a completely different
approach.

It has also been noticed that the upper limit of Landau quantum number
is a function of positional coordinate of the electron with which it is
associated within the WS cells. We have observed that at the surface
region, for all the values of magnetic field strength, this upper limit
becomes identically zero. Which actually means that the electrons near
the WS cell surface are strongly polarized in the opposite direction of 
external magnetic field. Whereas, for $B > 10^{15}$G, they are polarized
at every points within the cells.

It has been observed that the electron density within the cells
increases with the increase in magnetic field strength. Further, for all
the values of magnetic field strength, the electron density is maximum
near the nuclear surface ($r=r_n$) and minimum at the WS cell boundary
($r=r_s$).

The electron kinetic pressure is found to be positive near the central portion 
of WS cell but is negative near the surface region. There is a position within 
the cell at which the kinetic pressure vanishes. The position of this point 
changes with the strength of magnetic field. 
We have also studied the variations of kinetic energy, electron-nucleus
interaction energy, electron-electron direct potential energy and
electron-electron exchange interaction part within the cells. We have shown 
the variation
of these quantities within the WS cells for a number of magnetic field
strengths. 

In the second part of this work, we have investigated some of the
properties of inner crust matter of magnetars.
We have used HW and BBP equation of states for the nuclear mass formula.
It has been observed from both the models, that for a stable inner crust
matter, the nuclei present must be heavier than iron and much more
neutron rich. The heaviness is more in the case of BBP equation of
state. We have also noticed that for low and moderate values of magnetic
field strength, the variation of mass number and the corresponding
atomic number with magnetic field is not so significant. Whereas, for
$B \geq 10^{15}$G, when electrons occupy only the zeroth Landau
level, then much more heavier neutron rich nuclei are formed in the inner crust
region. It is found that high magnetic field behaves like a catalyst to 
generates heavy neutron rich nuclei.

We have observed that initially the electron density increases with the
increase in mass number, but as soon as free neutrons appear in the
system, the electron density decreases and saturates to some constant
value which depends on the magnetic field strength. In the case of HW
equation of state, free neutron density does not depend on the strength
of magnetic field, whereas, for BBP case, because of chemical
equilibrium condition, the free neutron density depends very weakly on
the magnetic field strength. We have noticed that in the case of BBP equation 
of state the overall qualitative difference is because of chemical
equilibrium among the constituents.

The total baryon density rises sharply like an avalanche for the value of $A$ 
at which free neutrons appear in the system. However, for BBP equation of state,
because of chemical equilibrium condition, the rise is not so sharply
visible for a given magnetic field $B$.

The qualitative nature of equation of states are almost identical. It is
found that in presence of strong magnetic field, the inner crust matter
becomes mechanically stable (with the positive value of kinetic
pressure) only at very high density.

We therefore believe that to study various properties of dense matter
associated with magnetars, one must consider all these significant
changes from the conventional scenario. Further, we expect that the
strong magnetic field, if present well within the inner crust of
magnetars, must affect the super-fluidity of cold neutron matter.

\newpage
\begin{table}
\caption{
Variation of the parameter $a$ and $b$ for $\phi(x)=ax+b$ and
scaled surface radius $x_s$ for WS cells,
parameters $\alpha$ and $\beta$ for $E_{KE}=\alpha x^\beta$
and the parameters $p$ and $q$ for $E_{ee}^{\rm{ex}}=p x^q$ with magnetic field 
strength}
\label{tab:1}

\begin{tabular} {|l|c|c|c|c|c|c|r|} \hline 
$B/B_c^{(e)}$ & $a$ & $b$ & $x_s$ (MeV$^{-1})$
& $\alpha$ & $\beta$ & $p$ & $q$ \\ \hline
$10^0$ & $-2.48$ & $1.0002$ & $30.67$ 
 & $0.0113$ & $2.694$ & $0.04$ & $0.574$\\
$10^1$ & $-2.54$ & $1.0001$ & $11.78$ 
 & $0.0789$ & $2.602$ & $0.25$ & $0.591$\\
$10^2$ & $-2.69$ & $1.0001$ & $4.58$ 
 & $0.7157$ & $2.504$ & $0.39$ & $0.743$\\
$10^3$ & $-3.08$ & $1.0002$ & $1.78$ 
 & $5.0198$ & $1.948$ & $0.52$ & $0.842$\\ \hline
\end{tabular}\par

\end{table}

\newpage
\begin{figure*}
\resizebox{0.75\textwidth}{!}{%
\includegraphics{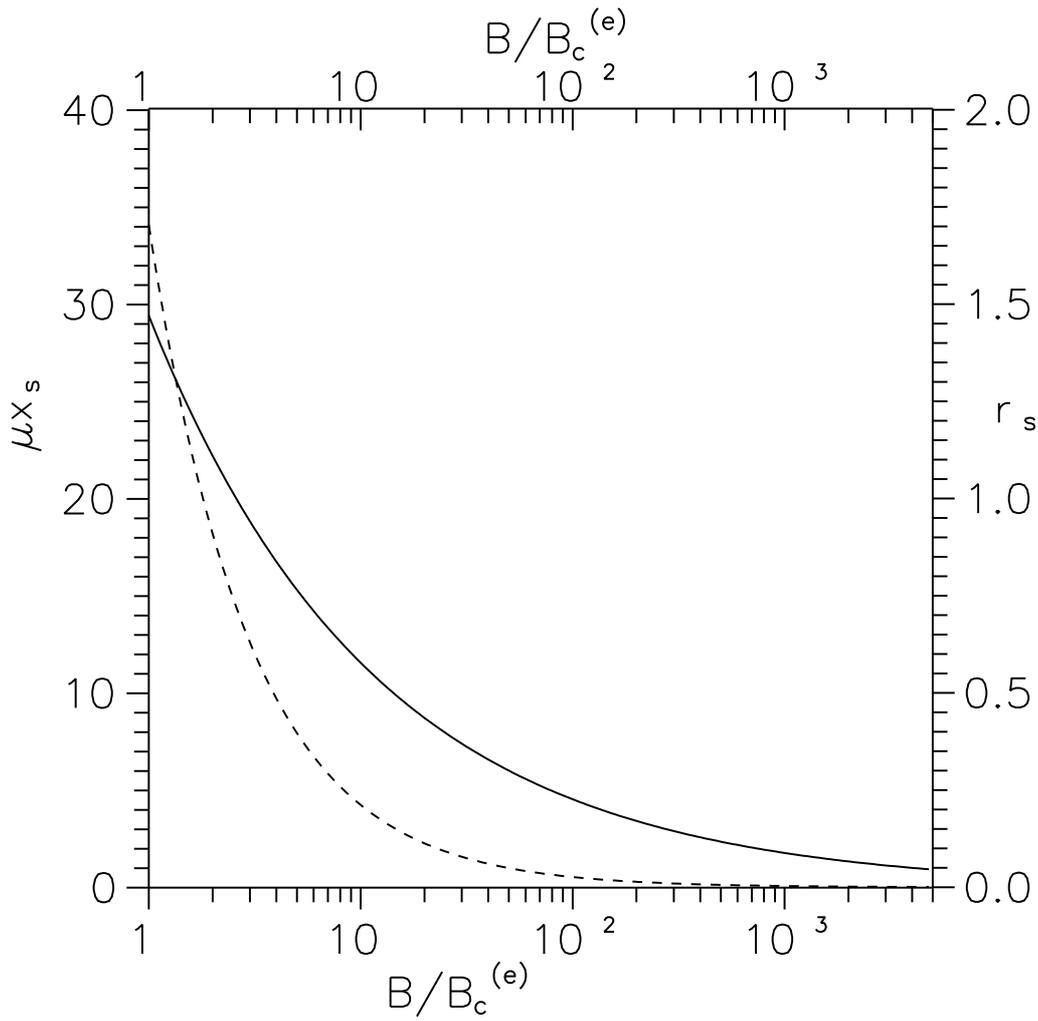}}
\caption{The variation of surface radius $\mu x_s$ in MeV$^{-1}$ (solid 
curve) and the actual radius $r_s$ in \AA unit
(dashed curve) with the strength of 
magnetic field (expressed in terms of $B_c^{(e)}$.)}
\label{fig:1} 
\end{figure*}
\begin{figure*}
\resizebox{0.75\textwidth}{!}{%
\includegraphics{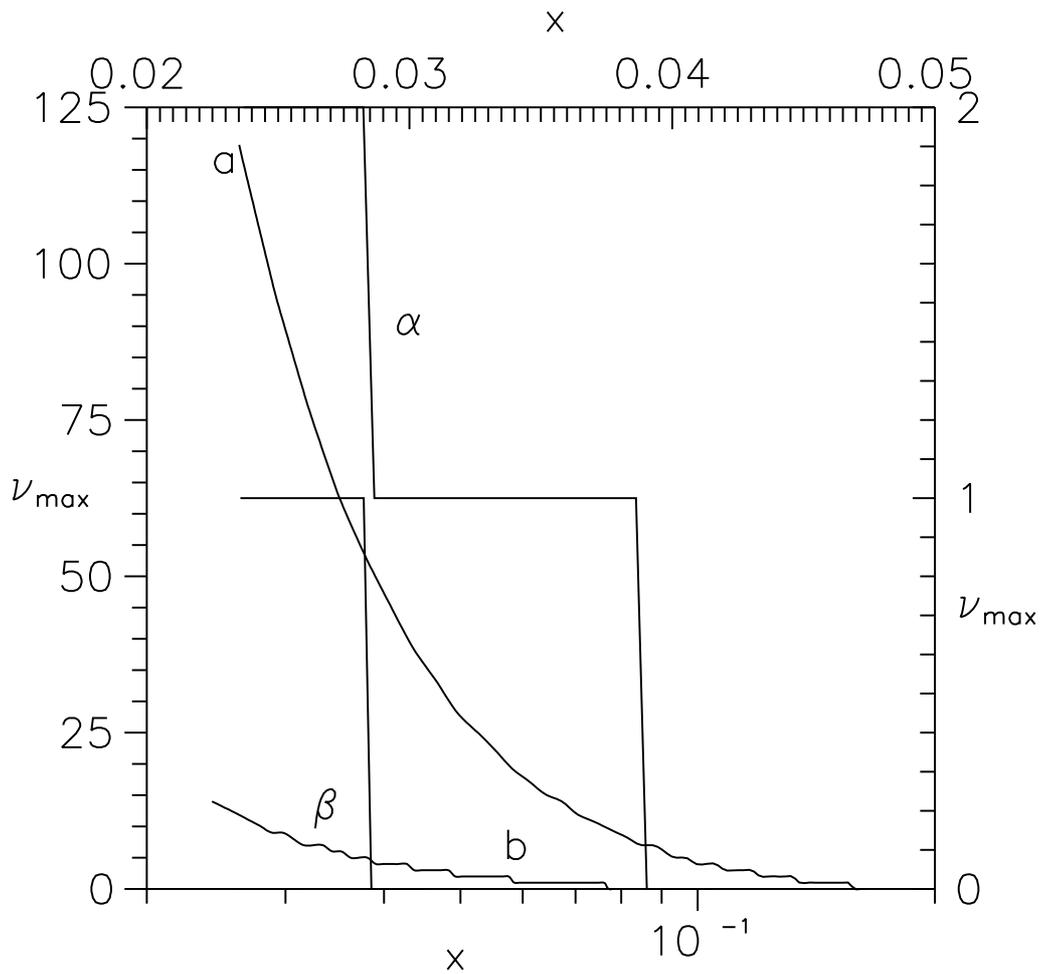}}
\caption{The variation of $\nu_{\rm{max}}$ with $x$ (dimensionless). For the 
curves indicated by the symbols $a$ and $b$, the relevant axes are at the 
bottom ($x$) and at the left
side ($\nu_{\rm{max}}$), whereas for the curves represented by the
symbols $\alpha$ and $\beta$, the 
axes are at the top and at the right side respectively.}
\label{fig:2} 
\end{figure*}
\begin{figure*}
\resizebox{0.75\textwidth}{!}{%
  \includegraphics{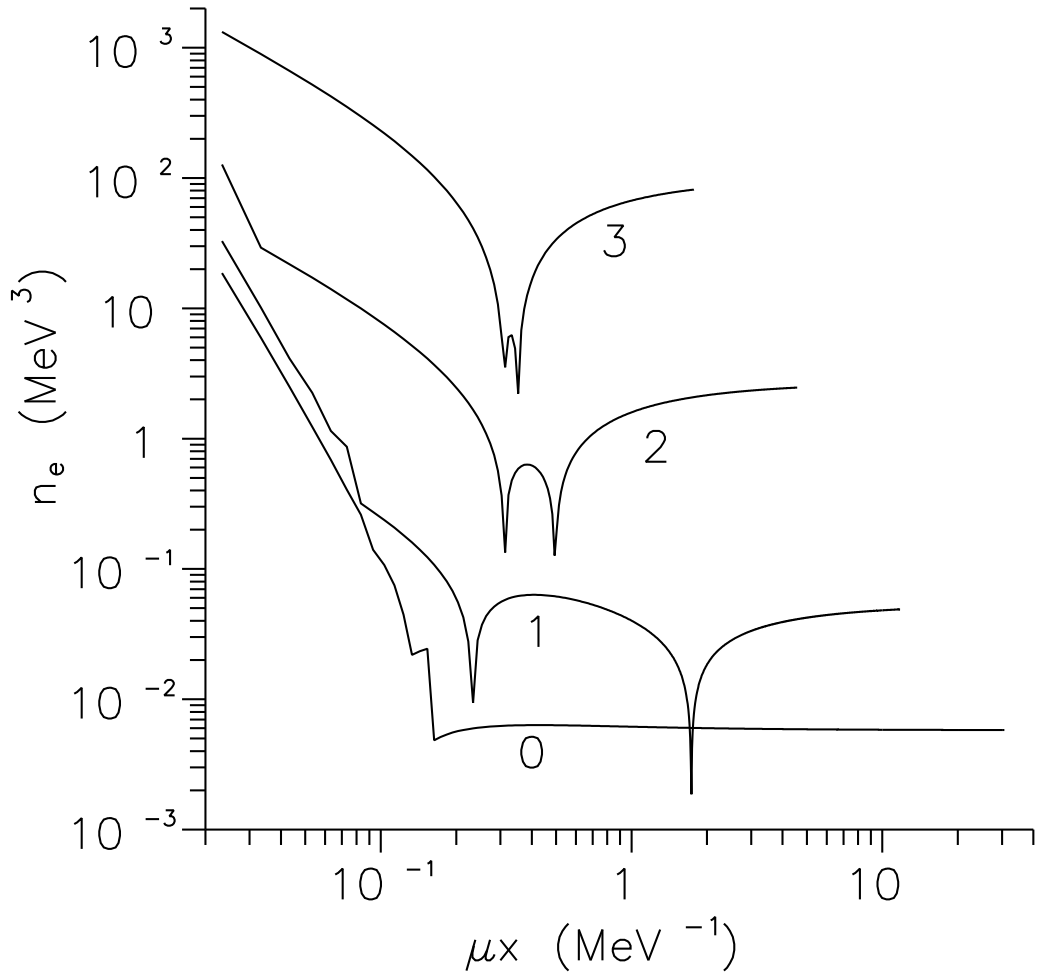}}
\caption{The variation of electron density in MeV$^3$ within the cells for
$B=h\times B_c^{(e)}$ with $h=1, 10, 100$ and $1000$ indicated by the
numbers $0$, $1$, $2$, and $3$ respectively. In order to get it in cm$^{-3}$, 
it is necessary to multiply each number on $y$-axis by $\approx 1.3\times
10^{32}$.}
\label{fig:3} 
\end{figure*}
\begin{figure*}
\resizebox{0.75\textwidth}{!}{%
\includegraphics{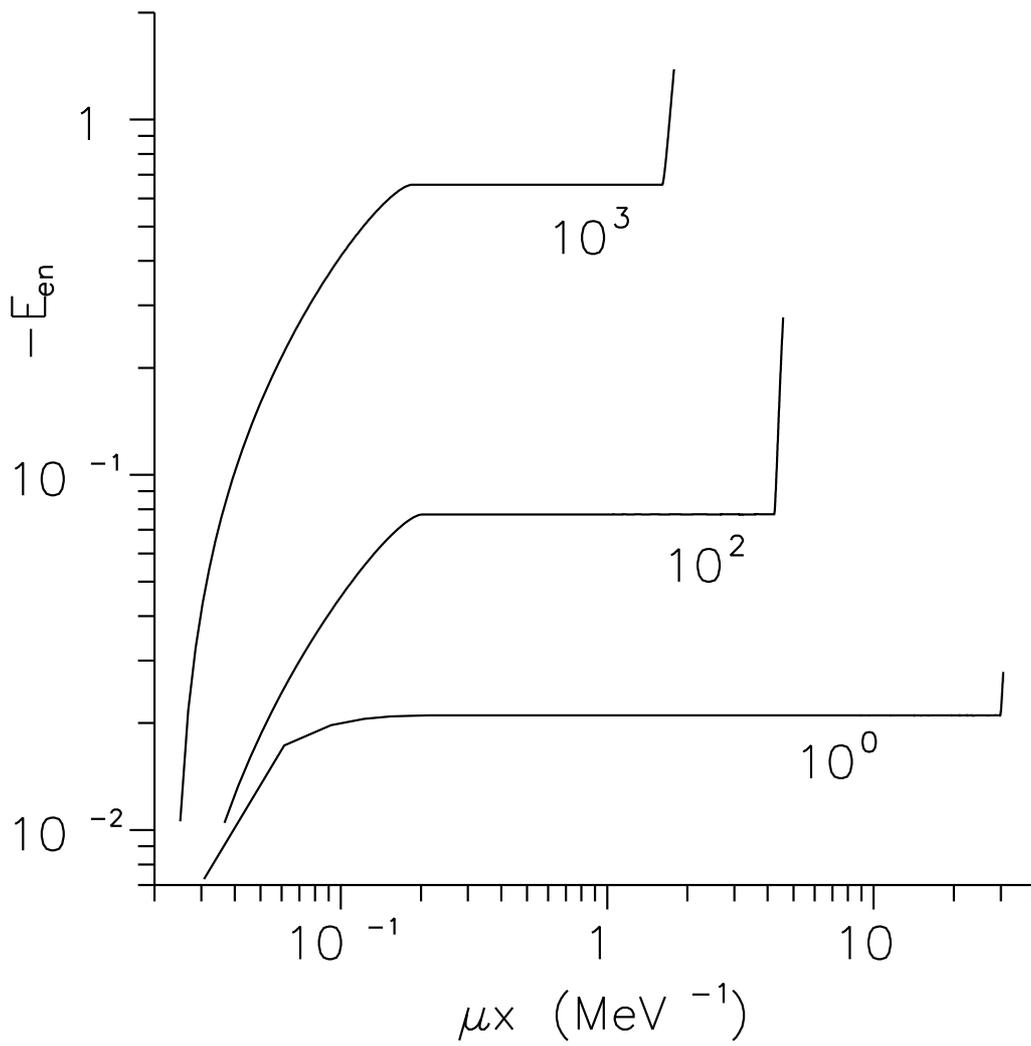}}
\caption{The variation of $-E_{en}(x)$, the electron-nucleus interaction energy
in MeV within the WS cells for $B=h\times B_c^{(e)}$ with $h=1, 100$ and
$1000$.}
\label{fig:4} 
\end{figure*}
\begin{figure*}
\resizebox{0.75\textwidth}{!}{%
  \includegraphics{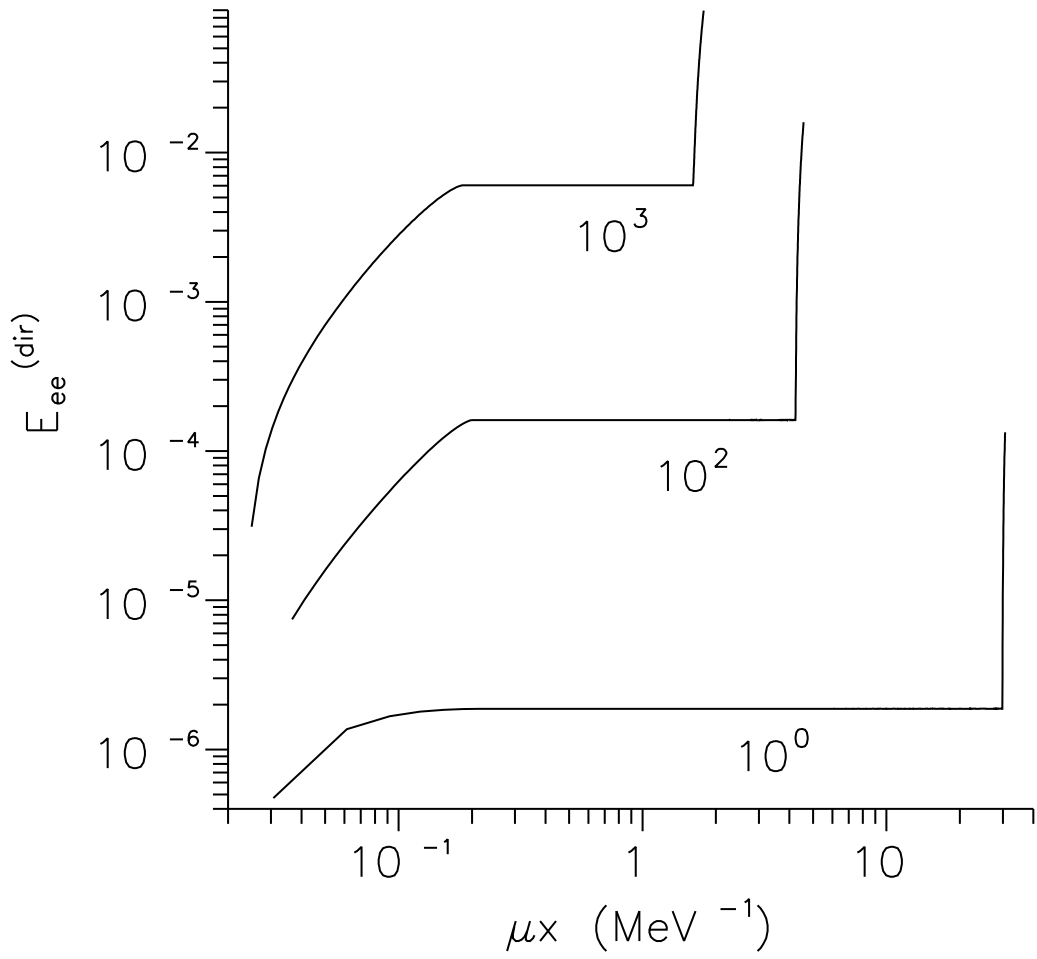}}
\caption{The variation of $E_{ee}(x)$, electron-electron direct interaction 
energy in MeV within 
the WS cells for $B=h\times B_c^{(e)}$ with $h=1, 100$ and $1000$.}
\label{fig:5} 
\end{figure*}
\begin{figure*}
\resizebox{0.75\textwidth}{!}{%
  \includegraphics{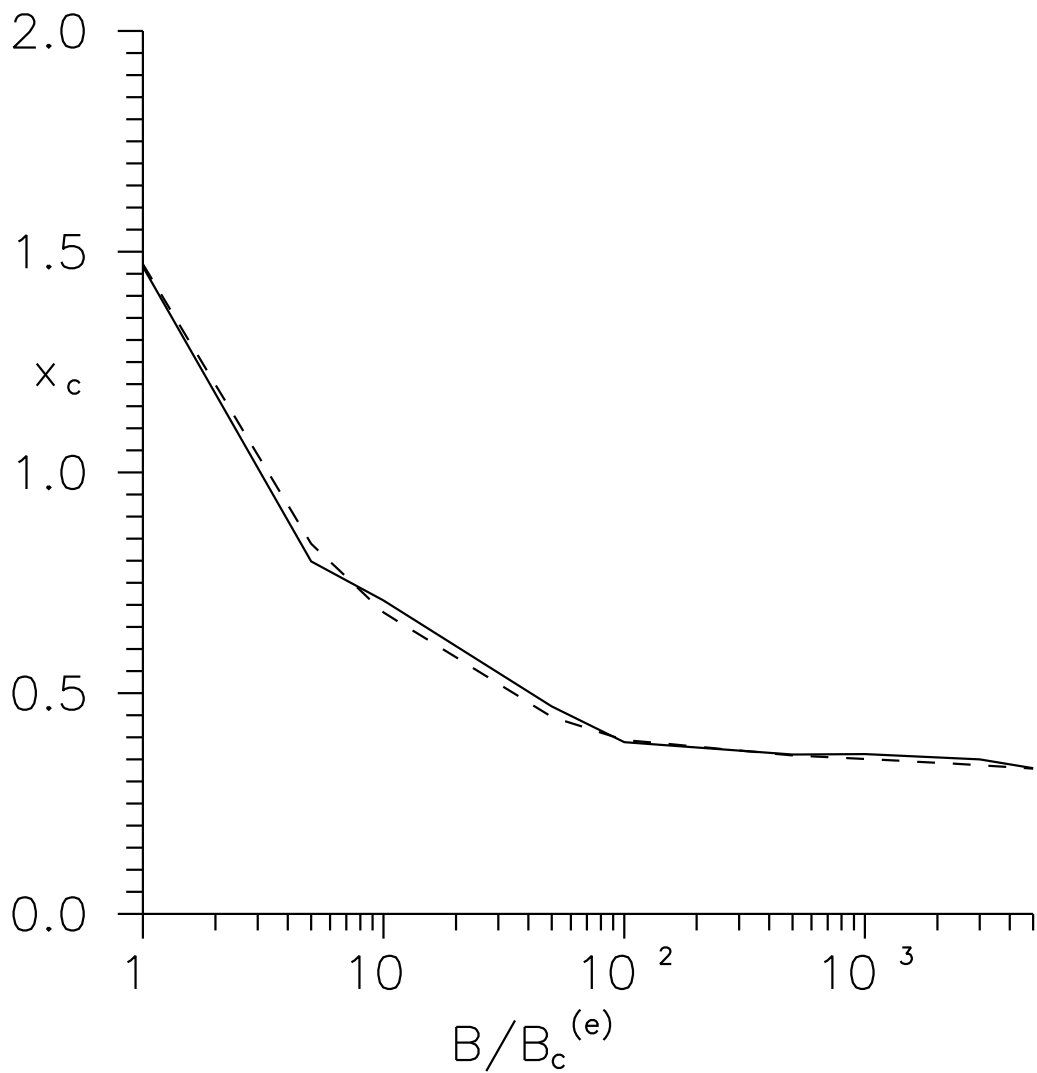}}
\caption{The magnetic field dependence of critical positions $x_c$
(which is dimensionless) within 
the cell at which the total kinetic pressure vanishes, solid curve 
is for the numerically evaluated points and the dashed curve is the fitted 
functional form.}
\label{fig:6} 
\end{figure*}
\begin{figure*}
\resizebox{0.75\textwidth}{!}{%
  \includegraphics{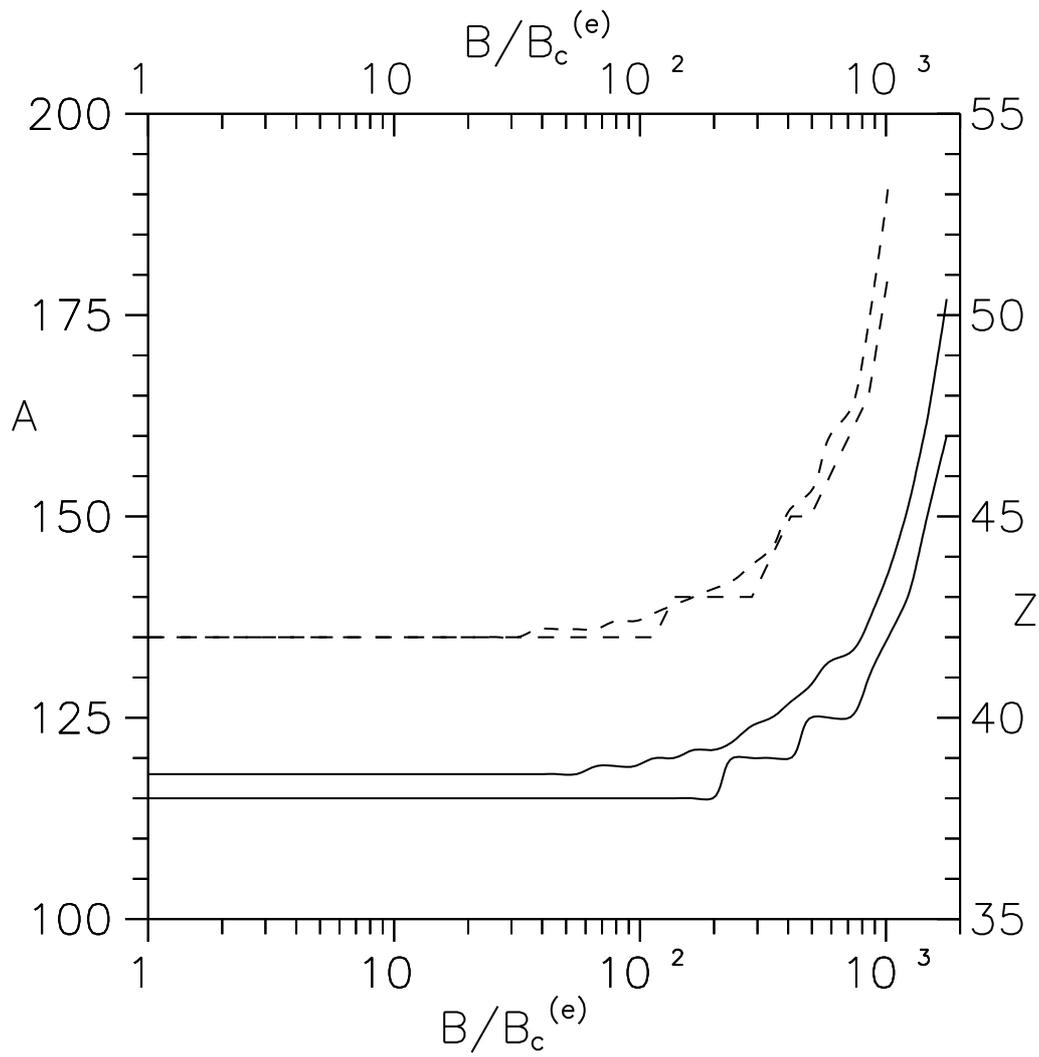}}
\caption{Variation of critical values for $A$ and $Z$ with the strength
of magnetic field $B$ (expressed as $B/B_c^{(e)}$) at which kinetic
pressure becomes just zero. The solid curves are based on HW equation of state
whereas the dashed curves are from BBP equation of state. In each
case, the upper curve is for $A$ and the lower one is for $Z$.}
\label{fig:7} 
\end{figure*}
\begin{figure*}
\resizebox{0.75\textwidth}{!}{%
  \includegraphics{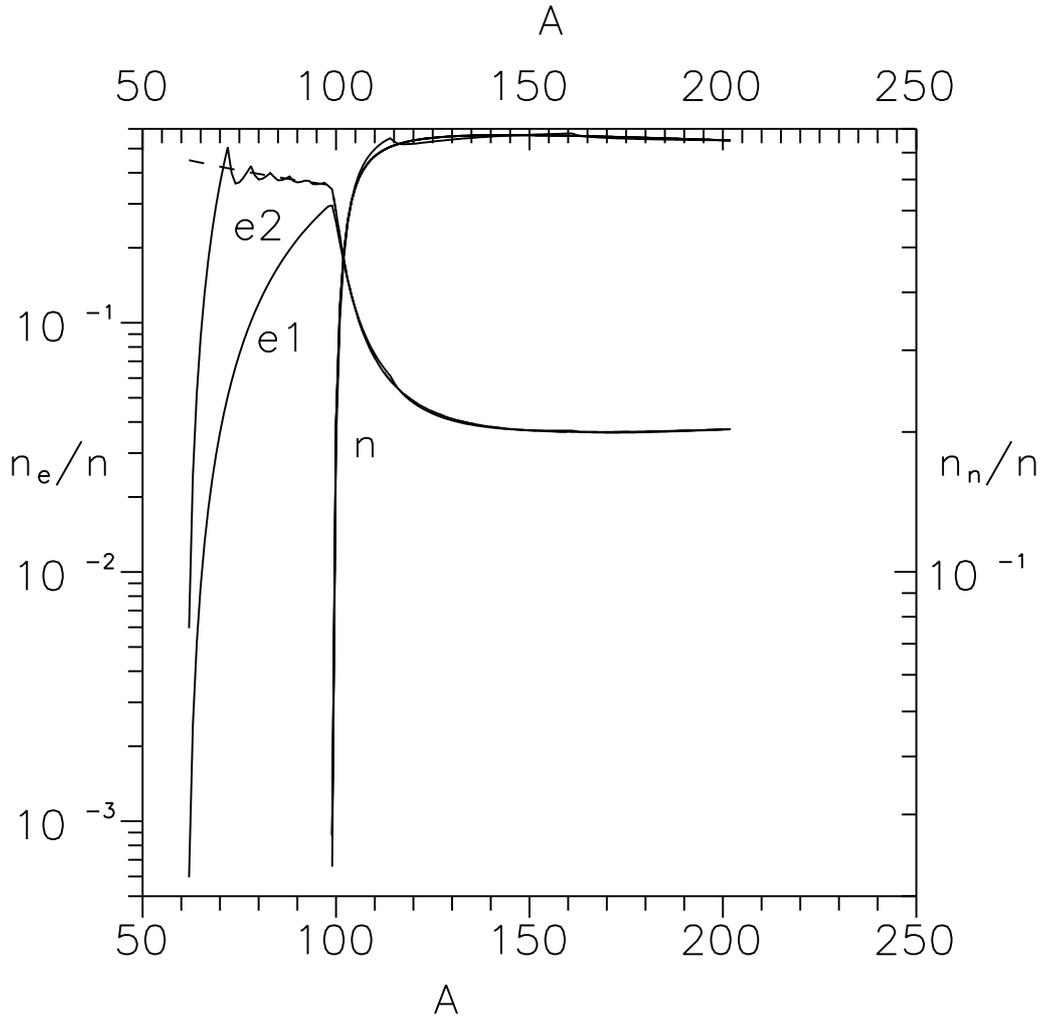}}
\caption{Variation of the ratio $n_e/n$ and
$n_n/n$ with the mass number using HW equation of state. Here $n_e$ is
the electron density, $n_n$ is the free neutron density and
$n=n_n+n_e A/Z$ is the total baryon density of inner crust matter.
Dashed curve is for $B=0$, middle and the lower curves are for
$B=10^2B_c^{(e)}$ and $10^3B_c^{(e)}$ indicated by $e1$ and $e2$ respectively. 
The curve indicated by $n$ is for $n_n/n$.}
\label{fig:8} 
\end{figure*}
\begin{figure*}
\resizebox{0.75\textwidth}{!}{%
  \includegraphics{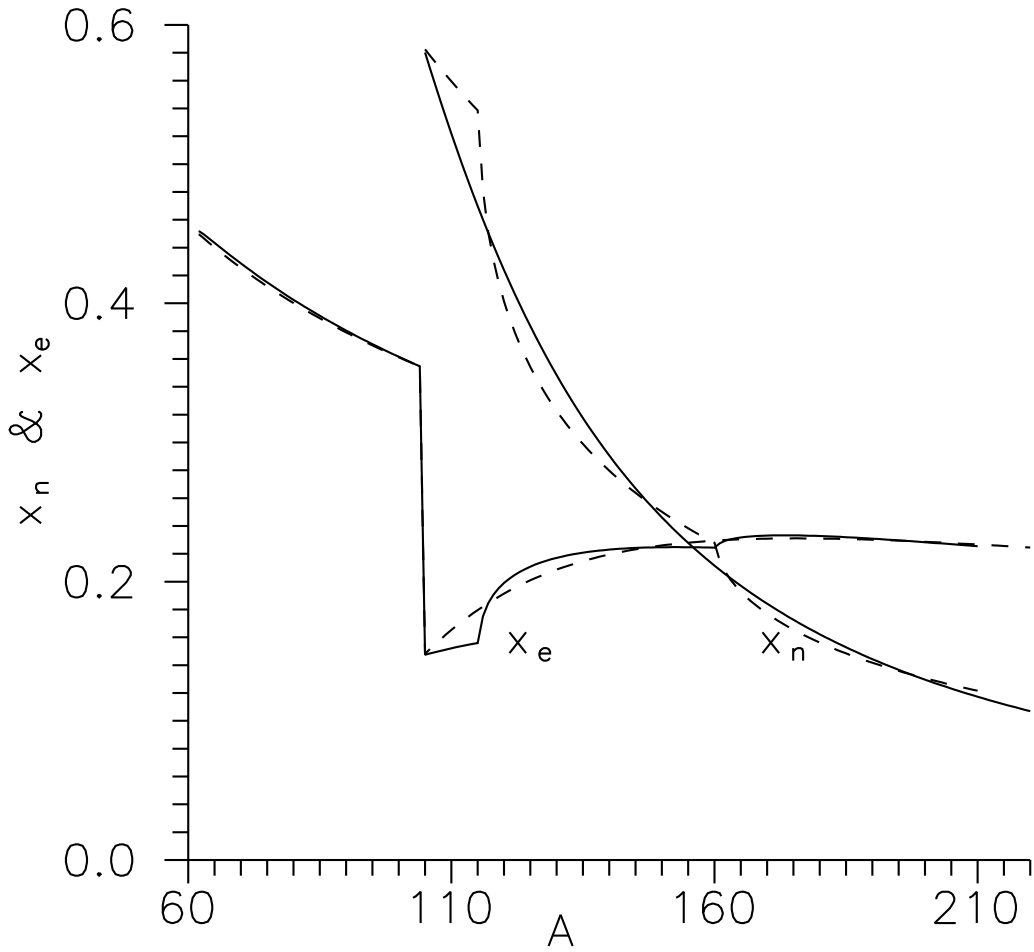}}
\caption{Variation of the ratio $x_e=n_e/n$ and
$x_n=n_n/n$ with the mass number $A$ using BBP equation of state. Here $n_e$ is
the electron density, $n_n$ is the free neutron density and
$n=n_n+n_e A/Z$ is the total baryon density of inner crust matter.
Solid curve is for $B=0$, the dashed curve is for
$10^3\times B_c^{(e)}$. Electron part and the neutron part are indicated by
$x_e$ and $x_n$ respectively.}
\label{fig:9} 
\end{figure*}
\begin{figure*}
\resizebox{0.75\textwidth}{!}{%
  \includegraphics{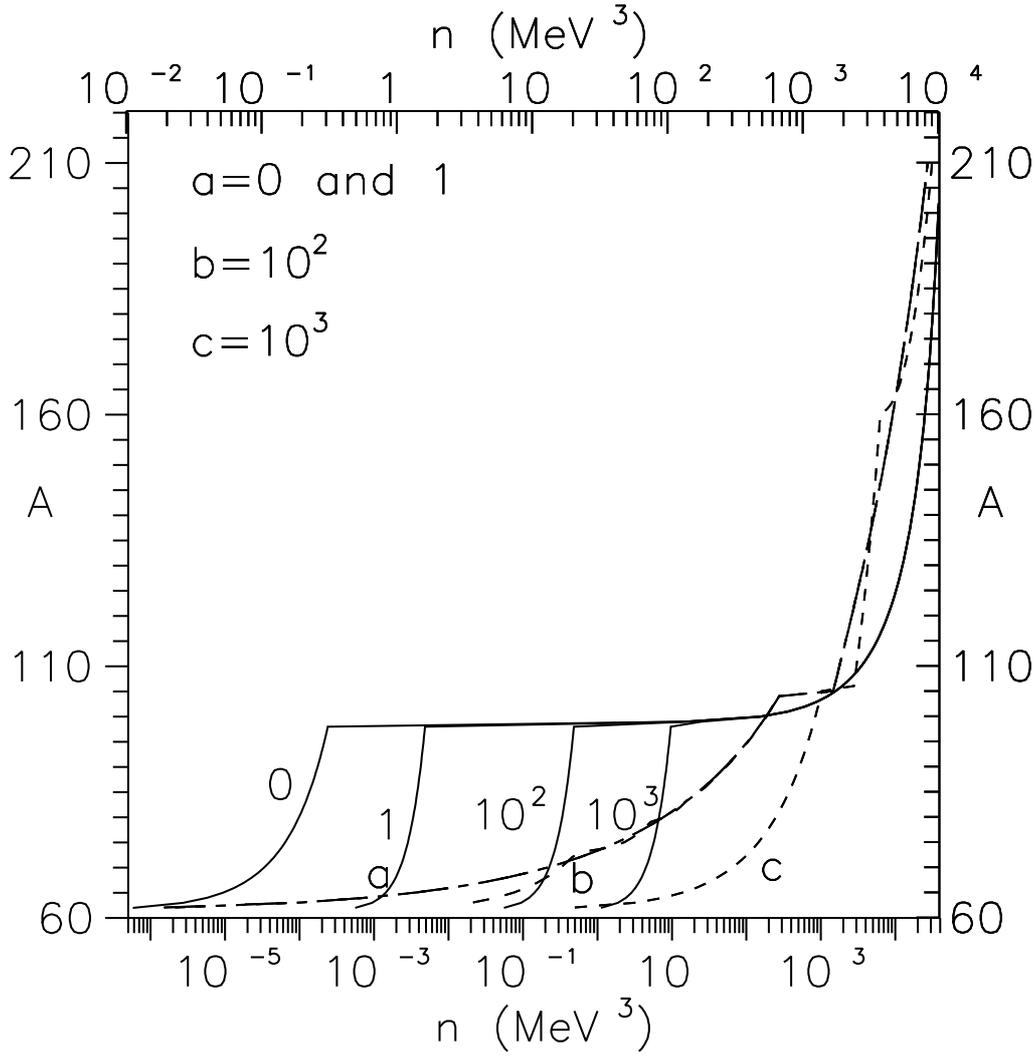}}
\caption{Variation of mass number of the nuclei with the total baryon density.
HW equation of state: four different magnetic field
strengths:$B=0$, $B=10B_c^{(e)}$, $B=10^2B_c^{(e)}$ and
$B=10^3B_c^{(e)}$, indicated by $0$, $1$, $10^2$ and $10^3$ respectively.
BBP equation of state: Dashed curves, the magnetic field strengths
$B=0$, $B=10B_c^{(e)}$ (both are indicated by $a$), $B=10^2B_c^{(e)}$ 
(indicated by $b$) and $B=10^3B_c^{(e)}$ (indicated by $c$). Here, the
lower x-axis is for HW equation of state and the upper one is for BBP
equation of state.}
\label{fig:10} 
\end{figure*}
\begin{figure*}
\resizebox{0.75\textwidth}{!}{%
  \includegraphics{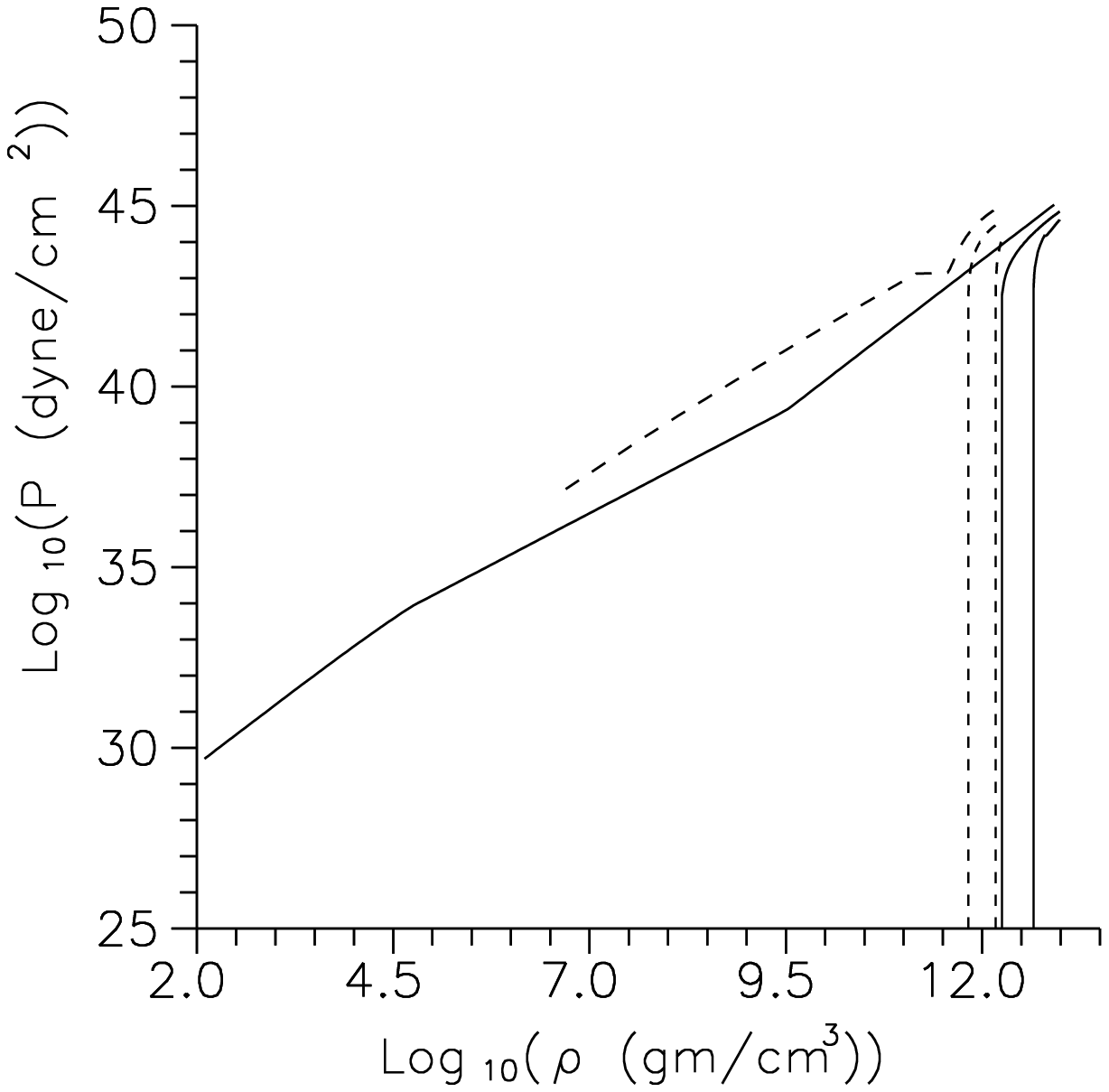}}
\caption{Equation of states from HW and BBP mass formulae. Solid curves are 
for HW equation of state, whereas, the dashed curves are for BBP case.
For each case, the upper curve is for $B=0$, middle and lower curves are for 
$B=B_c^{(e)}$ and $B=10^3B_c^{(e)}$ respectively. Please note that the
scales for $x$-axes at the top and bottom are different because of two
different equation of states.}
\label{fig:11} 
\end{figure*}

\begin{thebibliography}{}
\bibitem{R1} R.C. Duncan and C. Thompson, Astrophys. J. Lett. {\bf{392}},
(1992) L9; C. Thompson and R.C. Duncan, Astrophys. J. {\bf{408}},
(1993) 194; C. Thompson and R.C. Duncan, MNRAS {\bf{275}}, (1995) 255;
C. Thompson and R.C. Duncan, Astrophys. J. {\bf{473}}, (1996) 322.
\bibitem{R2} P.M. Woods et. al., Astrophys. J. Lett. {\bf{519}}, (1999) L139; 
C. Kouveliotou, et. al., Nature  {\bf{391}}, (1999) 235.
\bibitem{R3} K. Hurley, et. al., Astrophys. Jour. {\bf{442}}, (1999) L111.
\bibitem{R4} S. Mereghetti and L. Stella, Astrophys. Jour. {\bf{442}},
(1999) L17; J. van Paradihs, R.E. Taam and E.P.J. van den Heuvel,
Astron. Astrophys. {\bf{299}}, (1995) L41; S. Mereghetti,
to appear in the proceedings of Nato ASI Conference "The Neutron Star-Black 
Hole Connection", Elounda, Crete, June 07-18, 1999; A. Reisenegger, To appear 
in the proceedings of the conference on "Magnetic fields across
the Hertzprung-Russel diagram", ASP Conference, ed. G.Mathys, S.Solanki
and D.Wixkramasinghe, 2001; see also 
for current status on the observational aspects of magnetars,
S. Mereghetti, To appear in 44th. Rencontres de Moriond, La Thuile, on
"Very high energy phenomena in the universe", 2009.
\bibitem{R5} D. Bandopadhyaya, S. Chakrabarty, P. Dey
and S. Pal, Phys. Rev. {\bf{D58}}, (1998) 121301.
\bibitem{R6} S. Chakrabarty, D. Bandopadhyay and S. Pal, Phys. Rev.
Lett. {\bf{78}}, (1997) 2898;
D. Bandopadhyay, S. Chakrabarty and S. Pal, Phys. Rev.
Lett. {\bf{79}}, (1997) 2176.
\bibitem{R7} C.Y. Cardall, M. Prakash and J.M. Lattimer,
Astrophys. J. {\bf{554}}, (2001) 322 and references therein; 
L.B. Leinson and A. P\'{e}rez, astro-ph/9711216;
D.G. Yakovlev and A.D. Kaminkar, The Equation of States in
Astrophysics, eds. G. Chabrier and E. Schatzman P.214, Cambridge Univ;
S. Chakrabarty and P.K. Sahu, Phys. Rev. {\bf{D53}}, (1996) 4687;
S. Ghosh, S. Mandal and S. Chakrabarty, Ann. Phys.  {\bf{312}}, (2004) 398;
S. Mandal, R. Saha, S. Ghosh and S. Chakrabarty, 
Phys. Rev. {\bf{C74}}, (2006) 015801.
\bibitem{R8} S. Chakrabarty, Phys. Rev. {\bf{D51}}, (1995) 4591;
T. Ghosh and S.  Chakrabarty, Phys. Rev. {\bf{D63}}, (2001) 043006;
T. Ghosh and S. Chakrabarty, Int. Jour. Mod. Phys.  {\bf{D10}}, (2001) 89;
\bibitem{R9} Sutapa Ghosh, Sanchayita Ghosh, Kanupriya Goswami, 
Somenath Chakrabarty, Ashok Goyal, Int. Jour. Mod. Phys. {\bf{D11}}, (2002) 843;
Sutapa Ghosh and Somenath Chakrabarty, Mod. Phys. Lett. {\bf{A17}}, (2002) 2147.
\bibitem{R10} V.P. Gusynin, V.A. Miransky and I.A. Shovkovy, Phys. Rev.
{\bf{D52}}, (1995) 4747;
D.M. Gitman, S.D. Odintsov and Yu.I. Shil'nov, Phys. Rev.  {\bf{D54}}, (1996) 
2968; D.S. Lee, C.N. Leung and Y.J. Ng, Phys. Rev. {\bf{D55}},
(1997) 6504; V.P. Gusynin and I.A. Shovkovy, Phys. Rev. {\bf{D56}},
(1995) 5251.
\bibitem{R11} D.S. Lee, C.N. Leung and Y.J. Ng, Phys. Rev. {\bf{D57}},
(1998) 5224; V.P. Gusynin, V.A. Miransky and I.A. Shovkovy, Phys. Rev.
Lett. {\bf{83}}, (1999) 1291; E.V. Gorbar, Phys. Lett. {\bf{B491}}, 
(2000) 305; T. Inagaki, T. Muta and S.D. Odintsov, Prog. Theo. Phys. 
{\bf{127}}, (1997) 93, hep-th/9711084; T. Inagaki, S.D. Odintsov and Yu.I. 
Shil'nov, Int. J.  Mod. Phys. {\bf{A14}}, (1999) 481; 
T. Inagaki, D. Kimura and T. Murata, Prog. Theo. Phys. {\bf{111}}, 
(2004) 371; V.P. Gusynin, V.A. Miransky and I.A. Shovkovy, 
Phys. Lett. {\bf{B349}}, (1995) 477;
S.P. Klevansky, Rev. Mod. Phys. {\bf{64}}, (1992) 649;
C.N. Leung and S.-Y. Wang, Ann. Phys. {\bf{322}}, (2007) 701; Nucl.
Phys. {\bf{B747}} (2006) 266.
\bibitem{R12} Sutapa Ghosh, Soma Mandal and Somenath Chakrabarty, Phys. Rev. 
{\bf{C75}}, (2007) 015805.
\bibitem{R13} Nandini Nag, Sutapa Ghosh and Somenath Chakrabarty, 
Ann. Phys. {\bf{324}}, (2009) 499.
\bibitem{RR14} B.K. Harrison, M. Wakano and J.A. Wheeler, "Matter at
High Density; End Point of Thermonuclear Evolution", Brussels, Belgium,
(1958) pp. 124.; see also B.K. Harrison, K.S. Thorne, M. Wakano and J.A. 
Wheeler, Gravitation theory and gravitational collapse, University of Chicago
Press, Chicago, (1965).
\bibitem{R14} G. Baym, H.A. Bethe and C.J. Pethick, Nucl. Phys.
{\bf{A175}}, (1971) 225.
\bibitem{R15} see also G. Baym, C.J. Pethick and P. Sutherland,
Astrophys. Jour. {\bf{170}}, (1971) 299; F. Ferrini, Astr. Sp. Sci.
{\bf{32}}, (1975) 231; A.W. Steiner, Phys. Rev. {\bf{C77}}, (2008) 035805; J.
Margueron, N. Van Giai and N. Sandulescu, to appear in the proceedings
of EXOCT07, Catania, June 11-15, 2007; 
S.L. Shapiro and S.A. Teukolsky, Black Holes, White Dwarfs
and Neutron Stars, John Wiley and Sons, New York, (1983) pp. 42 and 188.
\bibitem{R16} E.H. Lieb and B. Simon, Phys. Rev. Lett. {\bf{31}}, (1973) 681;
E.H. Lieb, J.P. Solovej  and J. Yngvason, Phys. Rev. Lett.  {\bf{69}}, (1992) 
749; E.H. Lieb, Bull. Amer. Math. Soc., {\bf{22}}, (1990) 1;
\bibitem{MU12} R.O. Mueller, A.R.P. Rau and L. Spruch, Phys. Rev. Lett.
{\bf{26}}, (1971) 1136; A.R.P. Rau, R.O. Mueller and L. Spruch, Phys. Rev. 
{\bf{A11}}, (1975) 1865; S.H. Hill, P.J. Grout and N.H. March, Jour. Phys. 
{\bf{B18}}, (1985) 4665.
\bibitem{R17} R. Ruffini, "Exploring the Universe", a Festschrift in
honour of Riccardo Giacconi, Advance Series in Astrophysics and
Cosmology, World Scientific, Eds. H. Gursky, R. Rufini and L. Stella, Vol. 
{\bf{13}}, (2000) 383; Int. Jour. of Mod. Phys. {\bf{5}}, (1996) 507.
\bibitem{R18} R.P. Feynman, N. Metropolis and E.Teller, Phy. Rev.
{\bf{75}}, (1949) 1561.
\bibitem{R19} S. Chakrabarty, Phys.  Rev. {\bf{D54}}, (1996) 1306.
\bibitem{LY} Y.S. Leung, Physics of Dense Matter, World Scientific,
Singapore, (1984) pp. 35.
\bibitem{R20} T.D. Lee, Elementary particles and the universe: Eassays in
honous of M. Gellmann, ed. John. H. Schwarz, Cambridge University
Press, (1991) 135.
\end{thebibliography}
\end{document}